\documentclass[aps,amsmath,showpacs,twocolumn]{revtex4}
\usepackage{graphicx}

\begin{document}


\title{Cosmic Conspiracies}

\author{Douglas Scott} \email{docslugtoast@phas.ubc.ca}
\author{Ali Frolop} \email{afrolop@phas.ubc.ca}
\affiliation{Department of Physics \& Astronomy\\
University of British Columbia,
Vancouver, BC, V6T 1Z1  Canada}

\date{1st April 2006}

\begin{abstract}
The now standard vanilla-flavoured $\Lambda$CDM model has gained further
confirmation with the release of
the 3-year {\sl WMAP\/} data combined with several other cosmological
data-sets.  As the parameters of this standard model become
known with increasing precision, more of its bizarre features become
apparent.  Here we describe some of the strangest of these ostensible
coincidences.  In particular we appear to live (within $1\sigma$) at the
precise epoch when the age of the Universe multiplied by the Hubble
parameter $H_0 t_0=1$.
\end{abstract}

\pacs{01.50.Wg,02.50.-r,06.20.fa,06.30.Ft,31.15.Ar,99.10.Ln}

\maketitle

\date{today}

\noindent
The current cosmological picture has been remarkably successful,
but contains a number of apparently unnatural features.
This Standard Model of Cosmology \cite{SMC}
has received further support from the
analysis of the 3-year {\sl WMAP\/} data, combined with information from
other cosmological probes \cite{Spergel06}.  Some of the necessary
ingredients of the SMC have caused cosmologists to respond with adjectives
such as `epicyclic', `ugly', `baroque' or `preposterous', and with questions
such as `who ordered {\it that\/}?'

In particular, the Universe appears to be dominated by a form of `Dark Energy',
as well as containing considerably more Dark Matter than ordinary baryons.
Much is often made of the so-called `Cosmic Coincidence' problem, namely
that we live at the epoch where $\Omega_{\rm DM}\,{\sim}\,\Omega_{\rm DE}$,
despite the fact that the ratio varies dramatically with time \cite{Carroll}.
Another version of the same coincidence is that we live close to the epoch when
the Universe is transitioning from deceleration to acceleration.
But in practice these `coincidences' are no better than the factor
$\,{\sim}\,2$ level
(in fact $\Omega_{\rm DE}/\Omega_{\rm DM}\,{\simeq}\,3.8$ and
$z_{\rm jerk}\,{\simeq}\,0.8$), perhaps significant on an astronomical scale,
but hardly impressive in human terms \cite{bet}.

However, there is another version of this same `Dark Conspiracy'
that appears even more bizarre -- 
the current best-buy cosmology has values
of $H_0$ and $t_0$ which, within their uncertainties, are consistent with
$H_0 t_0=1$.  This is counter to the prevailing wisdom of the end of the
20th century, when it was expected that $H_0 t_0\simeq 2/3$, with unity
being considered the upper limit.

The new results from {\sl WMAP\/} data alone, specifically for a
flat $\Lambda$CDM model, give \cite{units}:
\begin{equation}
H_0 t_0 = 1.03 \pm 0.04.
\end{equation}
In other words, we seem to live exactly at (or at least within $1\sigma$ of)
the epoch when $t_0 \equiv H_0^{-1}$.
This would seem to be a conspiracy, designed to make the Universe
deceptively simple to explain!

When cosmologists teach introductory
material about the Universe, the following 2 concepts are often discussed:
(1) the overall curvature of the Universe could be negative, positive or
zero \cite{curvature}, {\it therefore\/}
we are trying to determine what it is; and
(2) if we run the clock backwards in time we end up with everything in the same
place at $t_0=H_0^{-1}$, {\it provided\/} that the expansion rate has been
constant.  But in today's precision cosmology reality,
we are forced to explain that the curvature appears to be
consistent with zero, so you can ignore that non-Euclidean stuff, and that
$t_0=H_0^{-1}$, even although the expansion rate slowed down for a while and
recently speeded up again! \cite{Misanthropy}

This value for $H_0 t_0$ is only natural in a completely empty universe,
otherwise known as the Milne model \cite{EAMilne}.
Even string theorists would typically agree that there is
empirical evidence for matter in the Universe.  Hence it seems unlikely
that the scale factor $a\propto t$, except as an average over
all cosmic time.  In fact, the current suite of
cosmological data requires not only $\Omega_{\rm M}>0$, but $\Omega_{\rm DE}>0$,
as well as independent constraints on 4 other parameter combinations --
a 6 parameter construction that has been
referred to as the new Milne model \cite{AAMilne}.

If this were the only odd factor in the SMC, then one might imagine that the
present-day Universe came about through sheer happenstance.  But there are
a large number of other apparent cosmological coincidences \cite{Anthropic},
not all of which
are stressed very often.  Could it be that, out of all the possibilities
in the Multiverse \cite{Multiverse}, we live in a peculiar universe,
with very precisely crafted properties?  Is it possible that
the Landscape is trying to tell us something?

A search for such conspicuous irregularities\cite{hunt} might turn up some of
the following:
\begin{itemize}

\item
$\Omega_{\rm DM} = 5 \Omega_{\rm B}$ : where does this value come from?

\item
$\Omega_{\rm \Lambda} = \pi \Omega_{\rm M}$ : somewhat increased
recently

\item
$\Omega_\nu\simeq\Omega_{\rm stars}$ : neutrino mass density similar
to that in stars

\item
$\Omega_{\rm M}\simeq Y_{\rm P}$ : relationship between matter and helium

\item
$t_0=3 t_\odot$ : 13.8 versus 4.6$\,$Gyr

\item
$t_0-t_{\rm jerk} \simeq t_\odot$ : Solar System formed at the Cosmic Jerk

\item
$z_{\rm B\gamma}\simeq z_{\rm lss}$ : baryon-photon equality at epoch
of last-scattering surface

\item
$h^2=1/2$ : dimensionless Hubble constant

\item
$\Delta^2_{\cal R} \sim n_{\rm B}/n_{\rm \gamma}$ : fluctuation power
comparable to baryon-to-photon ratio

\item
$\Omega_{\rm tot}=n=-w=1$ : the conspiracy of unities

\item
$T_0=$ triple point of water $\div\, 100$ \cite{whochosetemp}

\item
$\ell_{\rm peak} =$ Sunyaev-Zel'dovich null in GHz

\item
Stephen Hawking's initials in the {\sl WMAP\/} temperature image \cite{SH}

\item
The letters `C.O.' in the {\sl WMAP\/} synchrotron polarization pattern
\cite{Page06}

\end{itemize}

Which of these turn out to be complete coincidence remains to be seen.
But perhaps one of them will be the `smoking grail' that cosmologists have
been looking for to lead us beyond vanilla $\Lambda$CDM into a whole new
ice-cream parlour of models.



\smallskip



\baselineskip=1.6pt

\begin{thebibliography}{9}
\bibitem{SMC} Scott D., 2006, to appear in proceedings of `Theory Canada 1',
astro-ph/0510731.
\bibitem{Spergel06} Spergel D.N., et al., 2006, submitted to ApJ,
astro-ph/0603449.
\bibitem{Carroll} E.g.~Carroll S.M., 2001, Living Rev. Rel., 4, 1,
astro-ph/0004075.
\bibitem{bet} E.g.~`That was a close race, my horse only lapped your horse
once!' or `Amazing, your birthday's in October and mine is in July!'
\bibitem{units} Using the Markov Chain kindly provided by the {\sl WMAP\/}
team; it is also odd that in dimensionful
units the value is $1006\pm38\,{\rm km}\,{\rm s}^{-1}{\rm Mpc}^{-1}{\rm Gyr}$,
i.e.~very close to 1000.
\bibitem{curvature} Usually followed by much hand-waving, mumbling about
analogies with spheres whose surfaces you're trapped on and saddles which
don't really work in 3-dimensions anyway.
\bibitem{Misanthropy} Perhaps the Universe just hates us; this is known as
the Misanthropic Principle.
\bibitem{EAMilne} Milne E.A., 1934, Q.J. Math. Oxford, 5, 64; reprinted 2000,
GReGr, 32, 1949.
\bibitem{AAMilne} Milne A.A., 1927, `Now we are Six', Methuen, London;
reprinted 1999, Methuen, London; I am grateful to Max Tegmark for pointing
out this early cosmological treatise.
\bibitem{Anthropic} e.g.~Carter B., 1974, in `Confrontation of Cosmological
Theories with Observational Data', Reidel, Dordrecht, p.$\,$291;
Aguirre A., Tegmark M., astro-ph/0409072;
Rees M.J., astro-ph/0401424; Bousso R., Polchinski J., 2000, JHEP, 0006, 006;
Shemi-zadeh V.E., gr-gc/0206064; Martin B., 1998,
Skept. Inq., Vol.$\,$22, No.$\,$5, p.$\,$23.
\bibitem{Multiverse} Moorcock M., 1963, `The Blood Red Game', Sci. Fict. Adv.,
Vol.$\,$6, No.$\,$32, p.$\,$54.
\bibitem{hunt} In Scotland, such a search is called `hunt-the-gowk'.
\bibitem{whochosetemp} I thank Mark Halpern for pointing this out; see
also {\tt www.astro.ubc.ca/people/scott/whochosetemp.html}
\bibitem{SH} Which survive in the 3-year maps.
\bibitem{Page06} Page L., 2006, submitted to ApJ, astro-ph/0603450,
left panel of Fig.~8; suggestions for the significance of these initials
should be sent to the second author, Dr.~Frolop.
\end{thebibliography}
\end{document}